\documentclass[sn-basic]{sn-jnl}

\usepackage{etoolbox}
\makeatletter
\patchcmd{\ps@headings}
{\hbox to \hsize{\hfill Springer Nature 2021 \LaTeX\ template\hfill}}
{\hbox to \hsize{}}
{}
{}
\patchcmd{\ps@headings}
{\hbox to \hsize{\hfill Springer Nature 2021 \LaTeX\ template\hfill}}
{\hbox to \hsize{}}
{}
{}
\patchcmd{\ps@titlepage}
{\hbox to \hsize{\hfill Springer Nature 2021 \LaTeX\ template\hfill}}
{\hbox to \hsize{}}
{}
{}
\makeatother

\jyear{2021}%

\theoremstyle{thmstyleone}%
\theoremstyle{thmstyletwo}%

\theoremstyle{thmstylethree}%

\begin{document}

\title[NAS for Energy Efficient Always-on ML]{Neural Architecture Search for Energy Efficient Always-on Audio Machine Learning}


\author*[1]{\fnm{Daniel T.} \sur{Speckhard}}\email{speckhard@fhi.mpg.de}

\author[2]{\fnm{Karolis} \sur{Misiunas}}\email{kmisiunas@google.com}

\author[2]{\fnm{Sagi} \sur{Perel}}\email{sagipe@google.com}

\author[2]{\fnm{Tenghui} \sur{Zhu}}\email{ztenghui@google.com}

\author[1]{\fnm{Simon} \sur{Carlile}}\email{carlile.simon@gmail.com}

\author[2]{\fnm{Malcolm} \sur{Slaney}}\email{malcolm@ieee.org}

\affil*[1]{\orgname{X, The Moonshot Factory}, \orgaddress{\street{100 Mayfield Ave}, \city{Mountain View}, \postcode{94043}, \state{California}, \country{USA}}}

\affil[2]{\orgname{Google Research}, \orgaddress{\street{1600 Amphitheatre Parkway}, \city{Mountain View}, \postcode{94043}, \state{California}, \country{USA}}}


\abstract{Mobile and edge computing devices for always-on classification tasks require energy-efficient neural network architectures. 
In this paper we present several changes to neural architecture searches (NAS) that improve the chance of success in practical situations.
Our search simultaneously optimizes for network accuracy, energy efficiency and memory usage.
We benchmark the performance of our search on real hardware, but since running thousands of tests with real hardware is difficult we use a random forest model to roughly predict the energy usage of a candidate network.
We present a search strategy that uses both Bayesian and regularized evolutionary search with particle swarms, and employs early-stopping to reduce the computational burden.
Our search, evaluated on a sound-event classification dataset based upon AudioSet, results in an order of magnitude less energy per inference and a much smaller memory footprint than our baseline MobileNetV1/V2 implementations while slightly improving task accuracy. 
We also demonstrate how combining a 2D spectrogram with a convolution with many filters causes a computational bottleneck for audio classification and that alternative approaches reduce the computational burden but sacrifice task accuracy.}



%




\maketitle

\section{Introduction}\label{sec1}

It is challenging for the hearing impaired to identify important sounds such as running water, dogs barking, and crying babies. Typically, sound event classification systems feed spectrograms into image classification networks with great results~\citep{bib1}. Much of the sound-event classification work focuses on a paradigm where audio is sent over an Internet connection to a large neural network (such as ResNet50~\citep{bib2} with 20+ million parameters) that classifies the sound in the cloud. This approach relies on good internet data speeds. We instead search for a neural network that runs locally for a full day, continuously, on a battery powered device (e.g. smart watch, earphones, phone). This requires an energy efficient network to avoid draining the battery of the device prematurely and a small enough network to be able to fit into device memory.

Much of the platform aware neural architecture search (NAS) literature has focused on inference time (latency) as a user experience requirement for image classification~\citep{bib6, bib13}. Instead we think energy usage is the more important limiting factor.

An always-on audio model calculating an inference once every second makes 86400 inferences per day. As a result, the energy required per model inference is a critical matter when searching for the best architecture. A smartphone might have a battery capacity of around 51~kJ (e.g. Google Pixel 4~XL), a smartwatch around 3.6~kJ (e.g. Fitbit Versa 3) and earphones around 0.7~kJ, (e.g. Pixel Buds 2). For comparison, the baseline solution of deploying a high-performance network like MobileNetV2~\citep{bib3} on a Pixel 4~XL big core CPU uses 14~mJ per inference (1.21 kJ per day) when running sound event classification on spectrograms. It's evident that a network of that size will quickly consume a small device's battery capacity.

We introduce the first neural architecture search that incorporates the energy usage of the implementation. Our search also minimizes the memory usage of the neural network which, similar to energy usage, can be an equally limiting factor to model deployment on mobile and edge computing devices where total SRAM is limited. Our NAS builds upon related hardware constrained searches (e.g. searches constrained by hardware limitations such as memory). To find networks that also optimize for low energy and memory usage we need to incorporate these constraints into our reward function which we discuss in Section \ref{sec4}. 

We would like to guide our NAS with real hardware energy measurements. But at the scale we are operating (thousands of evaluations per task) this is prohibitive. In this work we train a random forest model on 10,000 candidate architectures from our search to accurately predict the energy usage of a candidate architecture which. We choose a random forest model since it random forests are known to work well on a variety of problem domains~\cite{bib7}. We also found that the random forest model outperforms a linear model. After running our search, we run the top three performing neural architectures on Pixel 4 XL CPUs five times to get average energy usage statistics.

We benchmark our work by comparing it to the efficiency of a state of the art network, in this case MobileNetV2, which performed well on the related task of 2D image classification (note, in this task we are classifying 2D spectrograms)~\cite{bib3}. Our NAS focuses on an audio classification task where the task and dataset are defined in Section \ref{sec5}. We constrain the maximum number (12) of sequential operations in a candidate neural network. For each sequential position in our candidate network, our search suggests an operation (e.g. either a 3x3 convolution or a 5x5 convolution). The possible operations in a NAS (described here as the search space) is often defined based on what operations are found in state of the art models on the task/dataset (in this case MobileNetV2). We discuss the search space more in Section \ref{sec6}.

Our search algorithm, evolutionary and Bayesian search algorithms defined in Section \ref{sec7}, suggests collections of block operations which define a candidate neural network. The search algorithm seeks to sugggest network architectures that optimize a reward function which scores each neural network candidate. We use early stopping, where we stop the training of unpromising architecture candidates to reduce the computational burden of the search.

In sum, we present a simple to implement neural architecture search that targets on-device energy efficiency, low memory usage for always-on audio models to satisfy the constraints outlined above. Our main contributions are: 

\begin{itemize}
 \item We introduce a multi-objective neural architecture search optimizes not only accuracy but also memory and energy usage. We employ both a Bayesian and evolutionary search algorithm with the evolutionary algorithm returning slightly better results.
 \item We train a random forest model to predict the energy usage of candidate neural network architectures in our search space. The model achieves a RMSE of 0.07 mJ per inference, which is a small fraction of typical energy usage per inference of our search space. This allows us to perform an architecture search that includes energy usage estimates without the added complication of including hardware in the search loop.Note, after the search we verify the performance of the winning implementations to make sure we have a real winner.
 
 \item We evaluate our method on a MobileNet based search space and find a model with accuracy slightly better than MobileNetV2 with 10x less energy usage, and 50x smaller memory footprint(Table~\ref{tab:results}).
 \item We show FLOPs are not a good proxy for energy usage even on a mobile CPU. Inference time (latency) is a better but imperfect proxy for energy usage. This is because power usage is not consistent across neural networks---memory access and arithmetic operations differ in power (Figure \ref{fig:flops_vs_energy}).
 \item Our search identifies a computational bottleneck created by combining spectrograms with 2D convolutional blocks (Table~\ref{tab:best_nas}) which is the typical architecture for audio classification. We show that an alternative approach of swapping the frequency axis with the depth axis of the spectrogram and using 1D convolutional blocks reduces energy usage, but significantly under-performs on the accuracy metric.
\end{itemize}

\section{Related Work}\label{sec2}

Several papers have explored neural architecture searches for neural networks intended for mobile devices. In particular, the MNAS paper of Tan [2019] performed their NAS where the inference time (latency) of architecture was included in the search reward function. Their search included a mobile phone in the search loop, where the candidate architectures ran on a mobile phone to return latency measurement.

In the TuNAS paper of Bender [2020], the authors avoided using hardware in the search loop to reduce software/hardware engineering requirements since it's significant work to connect mobile phones and measurement devices to the cloud where the NAS takes place. They instead opted to use a linear model to predict the inference time of neural network architectures in their search space. They train a linear model to predict the inference time of each architecture suggested by their search to rank each candidate architecture.

We, on the other hand, target energy per inference instead of inference time and also include a third term, memory usage in our reward function. Similar to TuNAS, we opt to avoid using hardware in the loop, and instead train a model (we use a random forest model instead of a linear model since it performs better) to predict the energy usage of each network architecture in our search space. 

TuNAS’ search algorithm creates a super (meta) network that includes all possible architectures into one network. It then drops out entire paths during training. This search algorithm is very efficient since only one (large) architecture needs to be trained instead of many possible candidate architectures trained separately. After training they then mask their network so that only a single path in the super network is active so they can score a single network's performance using the super network trained weights. However, for the trained weights of the super network to be similar to the weights of the lone architecture, significant paths in this network during training needed to be dropped out (turned off) which can make training unstable. The paper's authors also made this observation. This instability in training is the reason we instead opted to use Vizier’s algorithms instead which likely sacrifice computation time during the search.

MNAS and TuNAS use reinforcement learning (RL) to suggest new architecture candidates whereas we opt to use both a bayesian and genetic algorithm. We made this decision since NAS literature has shown evolutionary algorithms should yield similar results to RL for image classification tasks which we expect to behave similarly to our task of audio classification tasks on 2D spectrogram images~\cite{bib33}. The bayesian and genetic algorithms are also easier to setup out of the box.

Our search space is most similar to Wu's FBNet paper [2019]. The authors in this paper used a search space of different block operations. The search there was over the convolution kernel size, number of filters and expansion parameter of each block in the architecture. We similarly search over the kernel size and number of filters. However, FBNet only searches over MobileNetV2 like operations where we instead include MobileNetV1 blocks which hypothesize might be more energy efficient and other smaller block types that comprise a large block operation search space. Similar to FBNet, TuNAS also builds a MobileNetV2 based search space. We use a smaller maximum network size since we are targeting more energy efficient and memory efficient networks. 
We summarize our NAS search features and compare them to related hardware constrained searches in Table \ref{tab:best_nas}. One difference we don't include in the table to make it easier to visualize is that we focus on audio classification rather than image classification in our search.

\begin{table}
  \caption{Comparison of related hardware constrained neural architecture searches}
  \begin{tabular}{p{0.08\linewidth} | p{0.21\linewidth} | p{0.13\linewidth} | p{0.15\linewidth} | p{0.175\linewidth}}

    \toprule
    Search Name & Search Space & Constraints & Model for Constraints & Search Algorithm \\
    \midrule
    MNAS & MobileNetV2 based & Inference time & Hardware in the Loop & RL \\
    \hline & \\[-1.5ex]
    TuNAS & MobileNetV2 based & Inference time & Linear Model & Super Net + RL\\
    \hline & \\[-1.5ex]
    FBNet & MobileNetV2 based & Inference time & Latency Lookup Table & Super Net + RL\\
    \hline & \\[-1.5ex]
    This work & MobileNetV2\: + MobileNetV1 \: + Other blocks & Energy + memory use & Random \:\:\:\: forest & Evolutionary / Bayesian\\

  \bottomrule
\end{tabular}
\label{tab:best_nas}
\end{table}

\section{Optimization Criteria}\label{sec4}

We need to find a neural network that finds a balance between energy efficiency and memory usage, while still achieving state of the art accuracy. One option for our search would be to optimize accuracy while treating memory usage and energy usage as hard constraints. This yields equation \ref{eq:optimization_equation_hard_constraints} where $x$ is the evaluation dataset, ACC is the accuracy of a candidate network $h$ in our NAS search space H, $\text{MEM}$ is the memory footprint and $\text{ENERGY}$ is the energy usage per inference of the network.

\begin{equation}
\begin{aligned}
 \min_{h \in H} \quad & ACC(h(x))\\
 \textrm{s.t.} \quad & MEM(h) \leq M_{0}\\
 & ENERGY(h) \leq E_{0}    \\
\end{aligned}
\label{eq:optimization_equation_hard_constraints}
\end{equation}

As noted by the MNASNet authors this approach maximizes a single metric and does not yield multiple Pareto optimal curves~\citep{bib6}. We are looking for Pareto optimal models (e.g. models which have the maximum accuracy without increasing memory and energy usage). To approximate the Pareto optimal solutions, we combine these optimization constraints into a single objective via a weighted sum (note, MNASNet used a weighted product). In addition, we do not need the absolute lowest energy or memory and thus we limit the loss below an arbitrary threshold. The reward in Equation \ref{eq:optimization_conditions_single_equation} proportionally penalizes larger memory sizes and energy usages.

\begin{multline}
\: R = ACC(h(x)) \: \: -
b\,\max(0, \text{ENERGY}(h)-E_0) \\ -
c\,\max(0, \text{MEM}(h)-M_0)
\label{eq:optimization_conditions_single_equation}
\end{multline}

Memory and energy usage are penalized with a ReLU function that activates after the thresholds, $M_0$ and $E_0$, respectively, are crossed. In this study, we use an energy threshold, $E_0$, of 1.25 mJ per inference, which amounts to slightly more than 0.2\% of the Pixel 4~XL battery when the network is running one inference per second all day. Above the energy threshold, $E_0$, we explored two different slopes: b and b'. The harsher penalty $b$, is set to $\frac{0.02}{0.75 mJ}$. Thus, above this energy threshold a 0.75 mJ increase in energy per inference must give at least a 2\% increase in accuracy for the same reward. The less harsh penalty sets $b'=\frac{0.02}{1.75 mJ}$.

We use a memory size threshold, $M_0$, of 60~kB above which larger memory sizes are penalized with slope $c=\frac{0.02}{30 \text{kB}}$. The 60~kB threshold is chosen to allow the network to be deployed on a wide variety of SRAM limited devices (e.g. smartphones, smartwatches and earphones) which are expected to have several machine learning applications running simultaneously. The chosen slope means that a 90~kB model must have at least 0.02 accuracy points more than a 60~kB model for it to have a better reward. In the next five subsections, we discuss the quantized accuracy, measuring physical energy usage, approximate energy usage metrics, how we approximate energy usage during NAS using a random forest, and finally the memory usage in the reward function in Equation \ref{eq:optimization_conditions_single_equation}.

The following subsections give details of the quantized network (to avoid expensive floating point operations), energy approximations and memory usage.

\subsection{Quantized Accuracy}

\begin{figure}[ht]
    \centering
    \includegraphics[width=0.8\textwidth]{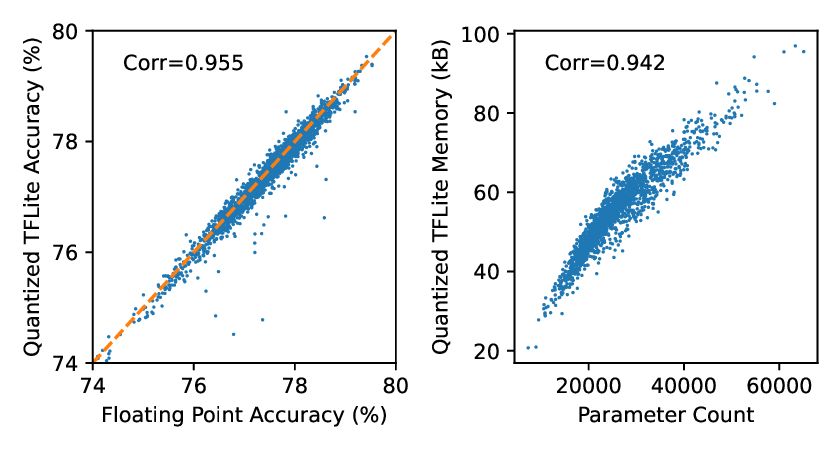}
    \caption{Quantized network performance. Left: Categorical accuracy versus int-8 quantized TFLite accuracy for two thousand candidate architectures sampled by the Vizier Bayesian (hybrid) algorithm. Right: The parameter count plotted against the quantized TFLite memory size}
    \label{fig:param_count}
\end{figure}

To measure the performance of the candidate architectures we use the accuracy of the 8-bit integer quantized TFLite model. The network is quantized using integer-8 quantization-aware training with the Tensorflow framework~\citep{bib23} to minimize the memory and energy usage. There is generally good agreement between the non-quantized accuracy and the quantized accuracy (correlation of 0.955) but there are some outliers (up to 6.5\% disagreement in accuracy) as seen in Figure~\ref{fig:param_count}. Since we are targeting on-device inference, we  use the quantized accuracy in our reward function.

\subsection{Physical Energy Measurements}

We use a Monsoon power monitor ~\citep{bib21} to measure the average power draw of a phone (without battery) running a candidate architecture.
The energy per network inference is platform dependent, thus for this paper we focus on the big core CPU of the Pixel 4~XL. During the measurement, we lock the CPU core frequency and use a single thread. The average inference time is measured using the TFLite benchmarking tool. We use these energy measurements in three ways: to check the approximations others have used (Section \ref{Energy_Approximations}), to train an approximate model to help guide the NAS (Section \ref{Approximating_Energy_using_Random_Forests}), and finally to verify the energy measurements shown in this paper (by repeating the measurement 5 times and reporting the mean and standard deviation).

\subsection{Inference Time and FLOPs as Energy Proxies}
\label{Energy_Approximations}

\begin{figure}[ht]
    \centering
    \includegraphics[width=0.8\textwidth]{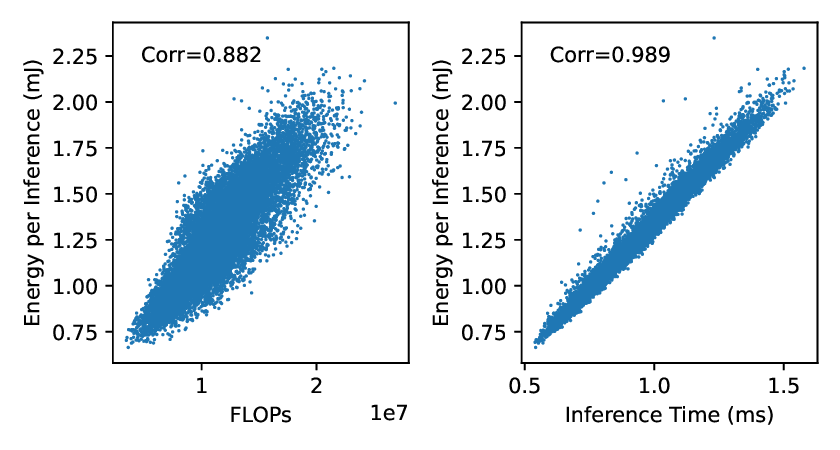}
    \caption{Two approximations to the total energy consumption using FLOPs (left) and inference time (right). These scatter plots are based on 15,000 randomly selected architectures in our search space measured on a Pixel 4~XL}
    \label{fig:flops_vs_energy}
\end{figure}

Other papers have used FLOPs (total number of floating point operations of the unquantized model) or inference time (latency) to approximate energy usage~\citep{bib14, bib18}. We discuss the drawbacks to these approaches. Figure~\ref{fig:flops_vs_energy} shows that a network with a FLOP count of 10 million might use between 0.6~mJ per inference to 1.5~mJ per inference. This agrees with what several authors have reported that the FLOPs count is a poor proxy for energy usage on-device, likely due to memory access not being accounted for in the FLOPs count~\citep{bib19}.

We find the correlation between the average inference time and the average energy per inference (which is simply the average inference time multiplied by the average power) is 0.989 over our search space (Figure~\ref{fig:flops_vs_energy}). Despite the good correlation, an average inference time of 0.85 ms could mean between 1.03 to 1.30 mJ per inference. This variation in energy usage is caused by variation in power draw between small architectural changes. We think these changes have an outsized influence to energy usage because of different parallelism and CPU cache optimisations. Thereby, in the same unit of time, the CPU may work to a different level of its full capacity due to different degrees of vectorized instructions. Note, the inference time of each network is also sometimes referred to as the latency of the model in computer vision NAS literature.

\subsection{Approximating Energy via a Random Forest for NAS}

In this paper, we have access to physical power measurements on Pixel devices. Measuring the energy usage of each NAS search candidate is a difficult software engineering task since the NAS search is happening in the cloud and would need to be connected to physical hardware that can automate the loading of the network, running of the network and average energy measurement. We instead opt to train a model that can accurately predict the energy usage of a given network. We then employ this model to estimate the energy usage of each NAS search candidate rather than getting physical measurements. This energy estimate is then fed into our reward function (Equation \ref{eq:optimization_conditions_single_equation}) so as to help us rank NAS network candidates. At the end of our search we gather the energy measurements of the best candidates on real hardware and report them in this paper.

Note, the alternative approach of using the inference time as an energy proxy would require us to either connect hardware in an automated way to our NAS search~\citep{bib6}, which as discussed is a difficult engineering problem or create a model for inference time \citep{bib13} to use during our search. We instead select a more direct route that avoids a complicated software/hardware engineering connection and use a model trained directly on energy usage data to predict energy usage.

We measure the average energy per inference of 15,000 architectures in our search space on the big core CPU of the Pixel 4~XL phone to train a model to predict energy usage for a given network. We employ a random forest (RF) model to predict the energy usage of models in our search space. We also tried a linear model, but it performed worse than the random forest at predicting energy usage of candidate architectures in our search space. The choice of a random forest model was motivated by the fact that decision trees are universal approximators (they can approximate any function) and random forests have been applied out the box to a wide variety of problems successfully \citep{bib7, bib8}. We used ten-fold cross-validation to tune the random forest hyperparameters.

\label{Approximating_Energy_using_Random_Forests}
\begin{figure}[ht]
    \centering
    \includegraphics[width=0.5\textwidth]{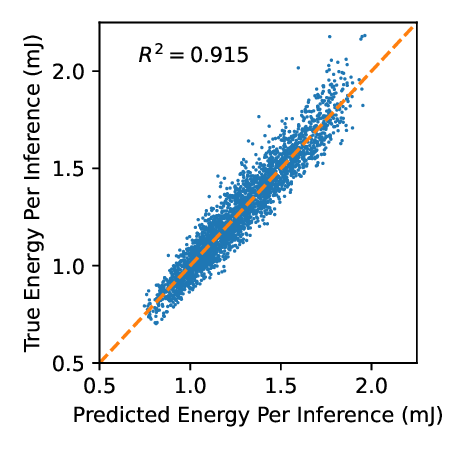}
    \caption{Random forest model fit to the energy per inference of networks in the search space}
    \label{fig:random_forest_energy}
\end{figure}

The random forest model takes as input the architecture parameters (e.g. kernel sizes/filter types of each block) as well as neural network level parameters, total FLOPs count and the TFLite memory size. The RF model has an $R^{2}=0.92$ and RMSE of 0.07 mJ per inference (Figure ~\ref{fig:random_forest_energy}). For reference, running a model with 1.3~mJ per inference from our search space five times on a pixel phone has an energy standard deviation (i.e. measurement noise) of 0.068 mJ. The RMSE values are close to the measurement noise from our phones, suggesting that the RF model is a good approximation for energy usage of a NN architecture. This allows us to perform NAS exploration fully on the servers, without remeasuring energy usage of each NN architecture permutation on the phone. We also tried running a linear regression model since the authors of TuNAS had success with a linear model~\cite{bib13}. The linear model achieved an $R^2$ coefficient of 0.89 and an RMSE of 0.089, both significantly worse than the RF model.

\subsection{Memory Footprint}

The SRAM available for small devices is somewhere between 10~kB to 1~MB and this memory is shared  with multiple applications. In this search we use the TFLite executable size, i.e. the static memory of the application in our objective. In Figure~\ref{fig:param_count} we compare the TFLite executable size to the parameter count of the network. We note that despite the parameter count being well correlated ($R^{2}$=0.94) to the quantized memory size of the network, it is still far from a perfect proxy. A parameter count of thirty thousand could mean anywhere between 50 to 65~kB of memory. The discrepancy is due to the integer-8 quantization-aware training with the Tensorflow framework~\citep{bib23} that we employ to minimize the memory usage.

\section{Sound Event Classification Dataset}\label{sec5}

We use the AudioSet dataset which contains over 2 million human-annotated 10 second sound clips derived from YouTube videos~\citep{bib20}. The AudioSet ontology contains more than 500 classes, but we use a subset of them to limit the complexity of our task. Specifically, we chose labels that mimic Sound Notifications on Android. The eight positive classes are (brackets indicate the original AudioSet labels, when multiple labels were mapped to one):

\begin{itemize}
 \item Alarms (fire alarm, smoke alarm, CO alarm)
 \item Baby crying
 \item Dog barking (dog, bark, yip, howl, bow-wow, growling)
 \item Door knocking
 \item Doorbell (doorbell, ding-dong)
 \item Phone ringing 
 \item Sirens (emergency vehicle, police car, ambulance, fire truck)
 \item Water running
\end{itemize}

We map all other classes in the AudioSet to a class labeled as the negative class. This tends to make this dataset somewhat challenging since the negative examples are all real sound events (e.g. guitar playing) and not simply low volume noise. In total we have 9 classes with one class being negative. We use the original train/evaluation/test split from AudioSet. We also ensure that our training/evaluation/test data is comprised of 50\% negative class examples. The log-mel spectrograms of the data are computed and augmented with SpecAugment~\citep{bib24}. We believe this mapping of AudioSet is a representative task for always-on sound event classification, while the dataset is also large enough for a NAS study.

\section{Search Space}\label{sec6}

\begin{figure}[ht]
    \centering
    \includegraphics[width=0.9\textwidth]{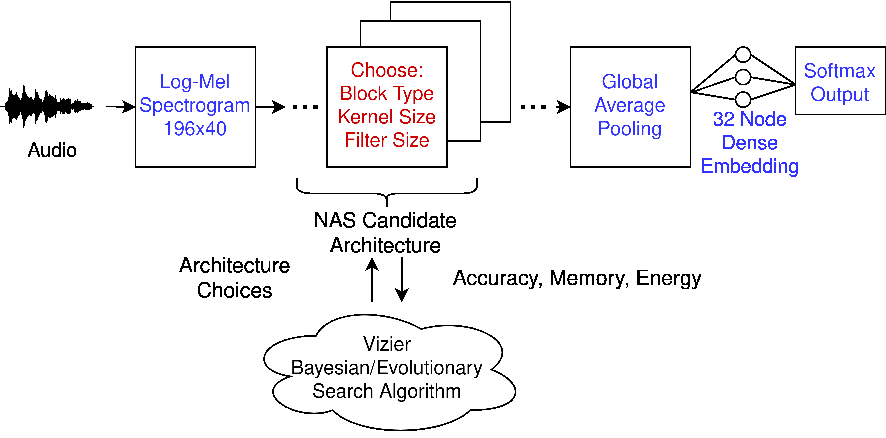}
    \caption{A block diagram of our network architecture search process}
    \label{fig:nas_search_diagram}
\end{figure}

The art of neural architecture searches lies in efficiently exploring a good search space. The search space defines the possible neural network architectures. A standard approach to define a search space is to first find a model that achieves good performance on the dataset and task of interest and decompose that model into its component blocks (e.g. a good performing network with a 5x5 convolution, 3x3 max-pool and 3x3 convolution with skip connection would decompose into a search space that includes these three operations)~\citep{bib11}. We make use of MobileNetV1~\citep{bib10} and MobileNetV2, which popularized depthwise separable convolutions, as benchmark models and use their two namesake block operations in our search space. We fix our network size to be twelve sequential blocks. We chose the number twelve after some (however not exhaustive) experimentation of how many blocks would be required to achieve MobileNetV2 like accuracy.

For every sequential block position in our network, we search for the block type, the number of output filters, and convolution kernel size. The search process is illustrated in Figure \ref{fig:nas_search_diagram}. Since the optimal parameters for each block are dependent on the position (e.g. a larger kernel is not so useful when the image size becomes very small towards the end of the network), we make the possible choices position dependent. Each of the twelve blocks has between nine to thirty options to choose from. In order to ensure the image size at the end of the network is the same for all possible candidates we fix the striding for all candidates to give a 7x5 image at the end of the network which is then fed into an average global pooling layer, before being fed into a constant 32 node dense layer that has 9 outputs (one for each class). We use a softmax activation on the logit outputs. The block macro-architecture is defined by the striding which is kept constant for each network and can be seen in Table~\ref{tab:best_nas}.

\subsection{KxK First Block}
The input to the first block is a spectrogram which has no depth dimension. This means using a more expressive block like a 2D convolution with a kernel size of KxK is computationally affordable. As such, like the MobileNet papers, we fix the first block type to be a KxK 2D Convolution, where K is the kernel size (i.e. an integer parameter over which the NAS algorithm should search). 

\subsection{Second Block}

The second block of our network is very important in terms of the computational load because the input image to this block is quite large since there has only been one block before to apply some striding to reduce the dimensionality. The first block acts on a 2D spectrogram. However, the second block acts on a three dimensional image (i.e. the input to this block now has depth) since the first block always applies more than one filter for all networks we consider in this work. As a result, we decide for this work to fix (hold constant) the second block's block type to be a Kx1 depthwise convolution followed by a 1xK depthwise convolution followed by a 1x1 pointwise convolution, which is the least computationally intense block in our search space. We call this block the Kx1-1xK-DW block (where DW stands for depthwise). Fixing the second block not only makes our search space smaller and thus more tractable to search, but returns a search space where almost 1/2 of the networks have less than our desired 1.25 uJ/inference energy usage target.

\subsection{Other Blocks}
The other ten blocks in our network use the block type choices of:

\begin{itemize}
 \item MobileNetV1
 \item MobileNetV2
 \item Kx1-1xK-DW
 \item MobileNetV2-Avg-Pool (only for stride (2, 2))
 \item Identity (only for stride (1, 1)).
 \item Kx1-1xK (only for last block).
\end{itemize}

The last block of the network has a small input size and as a result we also introduce the block choice of a Kx1 convolution followed by a 1xK convolution. When the striding of a block is (1, 1), we also add the choice of the identity block. This is done to ensure the output image is always the same size of every architecture.

\begin{figure}[ht]
    \centering
    \includegraphics[width=0.6\textwidth]{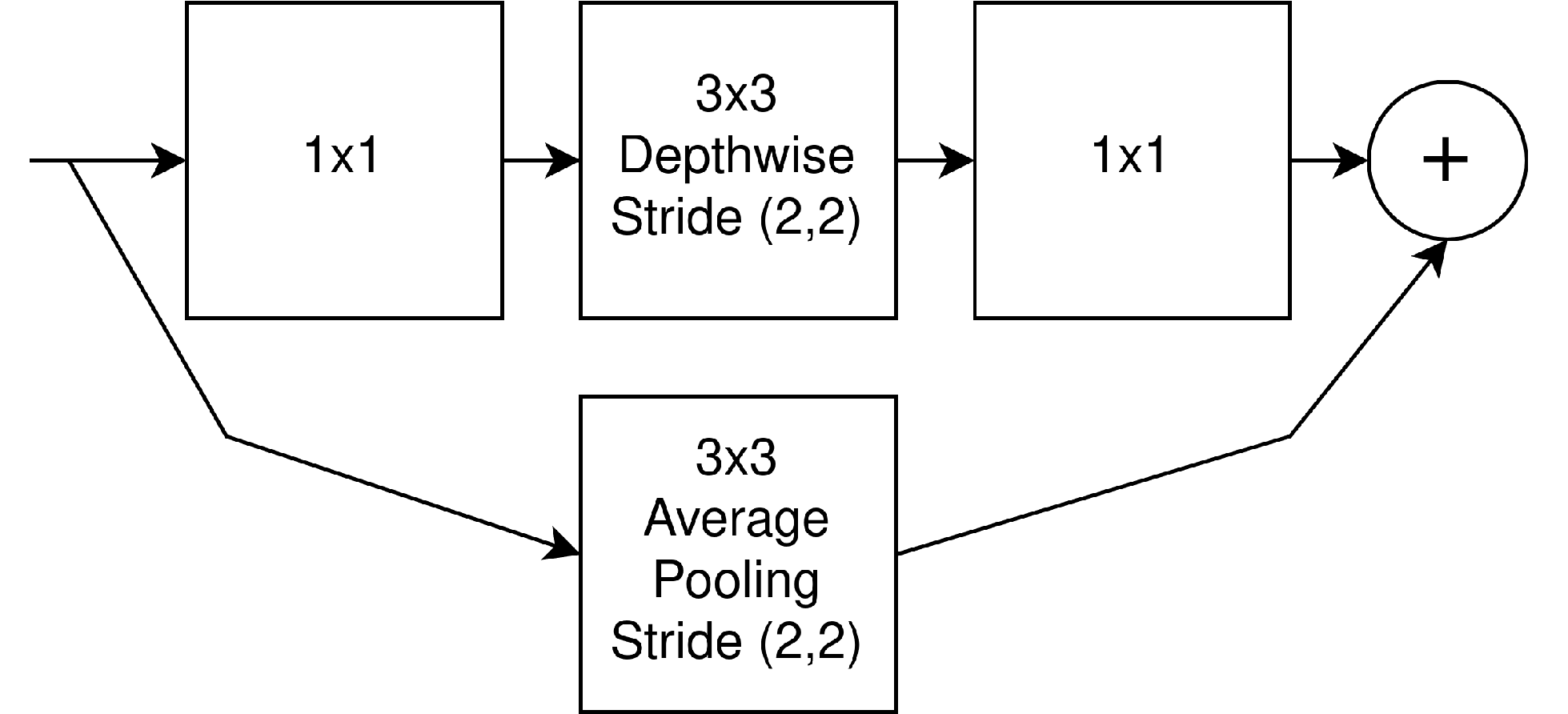}
    \caption{MobileNetV2 Block with an Average Pooling Parallel Path when strides are (2,2). This block variation is called MobileNetV2-Avg-Pool and is inspired by ShuffleNet}
    \label{fig:MobileNetV2_Avg_Pool}
\end{figure}

When using a striding of (2, 2), the original MobileNetV2 block does not contain a skip connection. The MobileNetV2 architecture is much larger than twelve blocks and most blocks in the original paper have a skip connection. Our network macro-architecture uses striding in five of the twelve blocks. We were motivated to add a parallel path to the MobileNetV2 block since we were worried information might be lost without it. The usage of parallel paths (residual/skip connections) on blocks was popularized and explained in the ResNet paper \citep{bib2}. We use a variation of the MobileNetV2 block, inspired by ShuffleNet we call MobileNetV2-AvgPool, so that when the striding is (2, 2) the input to the block takes a parallel path through a 3x3 average pooling layer with stride of (2, 2) as can be seen in Figure~\ref{fig:MobileNetV2_Avg_Pool}~\citep{bib16}.

We experimented with squeeze and excite blocks such as the ones in MobileNetV3~\citep{bib9}. We did not see any improvement in accuracy when adding MobileNetV3 blocks, only increases in memory footprint and energy usage. As a result, we left these blocks out of our search space.

\subsection{Kernel Sizes}

We search for the kernel size among \{3, 4, 5\} for the first five blocks. After the fifth block the input image is 13x10 and as a result we fix the kernel size for later blocks to be 3.

\subsection{Filter Sizes}

We choose among three options for filters for every block in the network, these choices are position dependent. We use filter sizes that are a multiple of 8 since we saw energy usage increase when using filter sizes that were not multiples of 8 on the Pixel 4~XL CPU. With these filter choices approximately one quarter of the architectures in our search space have memory footprints smaller than 60~kB and energy usages of less than 1.25~mJ per inference.

\subsection{1D Variant of the Search Space}

We also ran a modification of this search space that reduces all block types to their one dimensional counterparts (e.g. KxK convolutional kernel becomes a Kx1 kernel). The input spectrogram to the network is transformed by swapping the frequency axis with the depth axis as was done in the TCResNet paper~\citep{bib27}. This modification reduces the overall computational requirements, so we expect low energy usage but we were uncertain whether the one dimensional variant of our search space will return similar accuracy to MobileNetV1/V2.

\section{Vizier Search Algorithm}\label{sec7}

Our NAS is run on Vizier~\citep{bib25}, a black-box optimization service that removes much of the software engineering work necessary to efficiently run and analyze NAS runs. Our NAS trains each network separately and employs early stopping to eliminate architectures unlikely to contend for a top final objective value~\citep{bib14}. The alternative approach in NAS, training a single super-network with all architecture possibilities present (weight-sharing) saves computational resources but there are no guarantees the ranking of individual networks using shared weights are valid~\citep{bib12}. Our search space consists of fairly small networks that take one tenth of the time to train compared to a larger network like MobileNetV2. Note, the typical network in our search has between 15k-40k parameters (see Figure \ref{fig:param_count}), whereas MobileNetV2 uses more than 2000k parameters. As a result we did not explore other NAS search algorithms that train a single meta-network to avoid the expense of training candidate networks individually \citep{bib13}. We instead select a more computationally intense search by training each sampled network individually to three quarters of the full training time with some networks that appear unpromising stopped early.

Our search uses Vizier's algorithm to suggest candidate networks. Vizier suggests block types and their associated numbers of filters (filter size) for the twelve blocks in our network (note, two block types are fixed, the first and the second). For each suggested network, we calculate the memory footprint after converting the network to TFLite. We use a random forest Model described earlier to predict the energy usage of the candidate network. We then train the network for a fixed number of training steps or until Vizier determines the network is not likely to be a top candidate (early-stopping) while periodically evaluating the validation set accuracy. The best three candidates are re-trained with no-early stopping. Their accuracy on the test dataset is reported and their energy usage is evaluated ten times on a mobile phone with the mean result and standard deviation reported. After training each sampled architecture on the training dataset, we evaluate the reward on the evaluation dataset. The architectures with best rewards are retrained five times for 33\% longer on the same training set, and retested on the eval dataset. The best models from each NAS are retested with the unseen test dataset and these results are reported in this paper.

We employ two different search algorithms from Vizier for the NAS, one Bayesian and the other evolutionary, and run two thousand trials in each NAS experiment. Section~3 of Golovin's [2017] paper describes the Bayesian algorithm. The evolutionary Hyper-Firefly algorithm is an extension of the Firefly algorithm which uses regularization and particle swarms~\citep{bib26}. The Firefly hyperparameters are tuned by another Firefly algorithm every 50 iterations, using an objective metric equal to the best objective value over a sliding window of 50 iterations. 

Vizier's Bayesian algorithm slows down considerably (i.e. requires more time to produce a new suggestion) for our search space after a thousand trials. This is because a Gaussian process algorithm has $O(N^3)$ complexity, where N is the number of parameters multiplied by the number of trials. As a result, we switch from using the Gaussian process algorithm to the Hyper-Firefly algorithm after one thousand trials. Combining evolutionary algorithms with Bayesian approaches has been done before~\citep{bib12}. The results we obtain seem to generally favor the HyperFirefly (evolutionary) algorithm over the Bayesian (hybrid) algorithm. This could be due to the evolutionary algorithm being more explorative for the first one thousand trials.

\subsection{Early Stopping Algorithm}

Vizier can decide to stop training a network early, if it finds it unpromising. After Vizier suggests an architecture to train, the memory size and the predicted energy of the architecture are sent back to Vizier. On top of that information, the model in training is periodically evaluated and the intermediate evaluation accuracy is sent back to Vizier. If Vizier's early stopping model predicts that the current trial (architecture) will result in an objective worse than the best seen so far, with high confidence, the trial is stopped early. Early-stopping or performance curve stopping in Vizier is described in section 3.2 of Golovin's paper [2017]. This rule uses a Gaussian Process (GP) with a custom kernel to regress the evaluation curves of all available trials, where each input feature to the GP is a time bucket in the time series.

Temporal spatial stopping (TSGP) learns a Gaussian Process model for each time series, using the exponential curve kernel (\cite{bib29} Equation~6). The model also learns a mean function, at the asymptote, for each time series; and a mapping from the trial parameters to kernel parameters, allowing cross-trial information sharing.  This allows Vizier to make automated stopping predictions about each time series, which are informed by both a strong exponential prior, and the trial parameters.

We compared no early stopping, exponential decaying early stopping with default parameters, and exponential decaying early stopping with TSGP learned parameters. Experimentation with the three methods return very similar rewards. The TSGP early stopping used the least amount of computation (50\% less than forgoing early stopping) and as a result we employed it for our search.

\section{Results}\label{sec8}

\begin{table}

  \caption{NAS Results and Benchmarks. Top three models from each type of NAS are run five times against the test dataset and five times on a Pixel 4~XL CPU core for energy measurements. The bottom 4 rows show the baseline models.}
  \begin{minipage}{0.90\textwidth}

  \begin{tabular}{@{}p{0.26\textwidth} | p{0.10\textwidth} | p{0.09\textwidth} | p{0.10\textwidth} | p{0.10\textwidth} | p{0.09\textwidth} | p{0.10\textwidth}@{}}
    \toprule
    Model Name & Energy Per Inference~(mJ) & TFLite Memory Size (kB) & Accuracy (\%) & Inference Time (ms) & FLOPs & Param. Count\\
    \midrule
    NAS-Bayesian & $1.30 \pm 0.07$ & 51.22 &   $70.78 \pm 0.78$ & $0.98 \pm 0.00$ & 12.52M  & 23.95k \\
    \hline
    NAS-HyperFirefly & $1.27 \pm 0.07$ & 53.61 & $71.24 \pm 0.88$ & $0.94 \pm 0.00$ & 13.26M  &  24.94k \\
    \hline
    NAS-Bayesian-b' & $1.46 \pm 0.08$ & 56.96 & $70.89 \pm 0.50$ &  $1.07 \pm 0.01$ & 10.95M  &  27.13k \\
    \hline
    NAS-HyperFirefly-b' & $1.72 \pm 0.07$ & \textbf{45.10} & $\textbf{71.38} \pm \textbf{0.82}$ & $1.27 \pm 0.01$ & 15.61M  & \textbf{19.70k}\\
    \hline
    1D-NAS-Bayesian & $0.26 \pm 0.01$ & 55.38 & $67.04 \pm 0.64$ & $0.22 \pm 0.00$ & 6.43M  & 37.23k  \\
    \hline
    MobileNetV2 & $13.90 \pm 1.41$ & 2717.83 & $70.30 \pm 0.40$ & $0.99 \pm 0.04$ & 430.35M  & 2301.18k  \\
    \hline
    MobileNetV1 & $7.76 \pm 0.71$ & 3422.62 & $71.35 \pm 1.23$ & $5.47 \pm 0.01$ & 339.62M  &  3206.40k  \\
    \hline
    TCResNet8-1.0 & $\textbf{0.20} \pm \textbf{0.01}$ & 78.69 & $65.23 \pm 1.04$ & $\textbf{0.15} \pm \textbf{0.01}$ & \textbf{6.00M}  & 66.37k  \\
    \hline
    TCResNet14-1.5 & $0.64 \pm 0.06$ & 331.06 & $65.91 \pm 1.12$ & $0.43 \pm 0.00$ & 26.24M & 306.05k  \\
    \bottomrule
  \end{tabular}
  \end{minipage}
  \label{tab:results}
\end{table}

\begin{figure}[ht]
    \centering
    \includegraphics[width=0.8\textwidth]{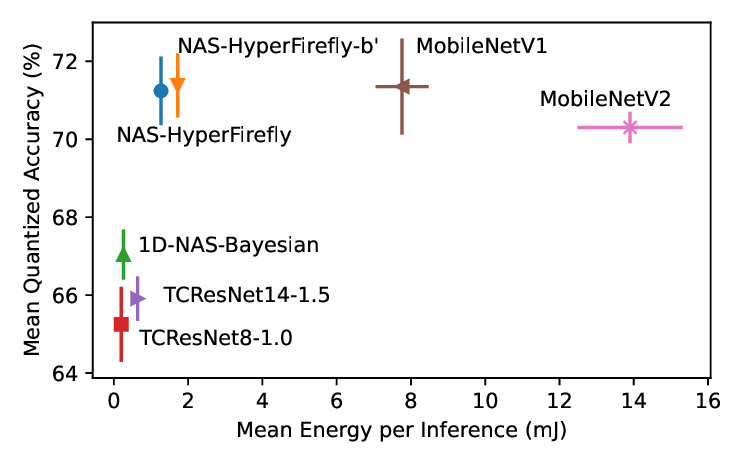}
    \caption{Mean energy per inference plotted against the mean quantized accuracy}
    \label{fig:visualize_table_results}
\end{figure}

Table~\ref{tab:results} conveys energy usage, memory size and accuracy for the best NAS results and two types of baseline models: MobileNet and TCResNet. We include MobileNetV1/V2 as baseline models since they are large models relative to our search space (i.e. they will have more capacity) and they have been applied successfully to audio classification~\citep{bib30} tasks. The MobileNetV2 benchmark uses an expansion parameter of six. At the other end of the model size spectrum, we include TCResNet models which are known to perform well on speech command recognition and require a low number of FLOPs and static memory. We use two different TCResNet model sizes, the TCResNet8 with width multiplier of one (labeled TCResNet8-1) and TCResNet14 with width multiplier of 1.5 (TCResNet14-1.5). Note, the benchmark models had to be slightly modified from their original paper version to work with a 196 by 40 size spectrogram input since MobileNets are designed to run on square input images and the TCResNet is designed to run on a 96x40 size spectrogram. The baseline models are all quantized with the same int-8 quantization-aware training used in the NAS architectures. Table~\ref{tab:results} reports all models' mean task accuracy after training five times to remove any bias from the initial starting condition. NAS done with the less harsh energy penalty $b'=\frac{0.02}{1.75}$ is marked with an accent suffix (b') in the Table. We visualize the NAS results from Table~\ref{tab:results} in Figure~\ref{fig:visualize_table_results}.

NAS-HyperFirefly, the best network when using the harsh energy penalty (note the HyperFirefly suffix means the HyperFirefly regularized evolution with particle swarm algorithm was used), achieves slightly worse accuracy than MobileNetV1 but better accuracy than MobileNetV2. Compared to MobileNetV2 it uses 50x less memory usage and 10x less energy usage. NAS-HyperFirefly-b', which was run with a less harsh energy penalty than NAS-HyperFirefly, uses more energy than NAS-HyperFirefly but also achieves better accuracy. NAS-HyperFirefly-b' achieves slightly better mean accuracy than MobileNetV1, the baseline model with the best mean task accuracy.

The networks using one dimensional convolutions, as was done in TCResNet, in Table~\ref{tab:results} tend to use very little energy but all return poor accuracy. 1D-NAS-Bayesian is the best NAS result when using only one dimensional convolutions. It achieves significantly better accuracy than both TCResNet baselines but with slightly more energy usage than TCResNet8-1.0.

The mean energy per inferences measured on the Pixel 4~XL CPU are not far from the predictions of the random forest model used during the NAS search. For example, the NAS-HyperFirefly uses 1.27~mJ per inference on average where the RF prediction used by Vizier during the NAS was 1.35~mJ. Similarly, the NAS-HyperFirefly-b' uses on average 1.72~mJ per inference and the RF prediction was 1.69~mJ.

\section{Discussion}

Table~\ref{tab:best_nas} shows the best performing network structure found by NAS-Bayesian. Of interest is that the network uses the kernel size 5 twice in the network---the larger receptive field must allow the network to improve the task accuracy despite costing more computationally. The network uses both MobileNetV1 and V2 blocks and the new MobileNetV2-avg-pool and 1xK-Kx1-DW block we introduce in this paper. This block type heterogeneity agrees with what many NAS authors have found that it can be beneficial to have different types of block structures~\citep{bib28}. It also shows MobileNetV2 blocks are not always superior to MobileNetV1 blocks when accuracy, energy and memory usage are all taken into account.

The number of output filters in the first block of the network creates a computational bottleneck when using 2D convolutions on a spectrogram. Table~\ref{tab:best_nas} shows the FLOPs of each block in the optimum NAS-Bayesian network and we see the second block uses 2M FLOPs. This is despite the block having 8 input filters and using a kernel size of 3 and 24 output filters. If the number of input filters to the second block were instead 24, the FLOPs count would triple to 6M.

The computational burden of the number of output filters in the first block is also seen in the energy usage of the network. The two most important features of our RF model that predicts energy are: the number of filters in the first block (59\%) and total number of FLOPs in the network (30\%). The rest of the features had 2\% or less impact. This shows the importance of the number of output filters used by the first block which creates a computational bottleneck in the network when using a spectrogram input to the network.

MobileNetV1 performs better than MobileNetV2 (expansion of~6) on this dataset. One of the main differences MobileNetV1 has to MobileNetV2 is a 1000 node dense embedding layer at the end of the network. The lack of the embedding layer may partly explain the poorer performance of MobileNetV2 which uses 700 kB less memory than MobileNetV1.

We note the poor accuracy performance of the 1D NAS variants was to be expected since the input image to the second block is now a 2D image (no longer 3D). However, the 1D-NAS-Bayesian using 0.26 mJ per inference compared to 1.3 mJ for the NAS-NAS-Bayesian model. For some applications, such as those deployed on batteries smaller than a mobile phone (earphones, smart watches) we can envision the 1D model being favored for using five times as little energy per inference. For such low-power use cases, we suggest further research into expanding the search space to use 1D convolutional blocks and/or combining this approach with model compression techniques (e.g. weight pruning).

Our search took roughly 15,000 GPU hours for the 2D NAS search sampling 2000 candidate networks. We used about one fifth that time for the 1D NAS variants. We did not experiment with other NAS methods to reduce the search's computational burden, which is something we would like to explore in the future. 

The NAS approach presented in this paper succeeds in finding a model (NAS-HyperFirefly) that is 10x more energy efficient and gives an improvement in absolute mean accuracy of 0.94\% compared to MobileNetV2. In comparison to MobileNetV1, we find a 4x more energy efficient network (NAS-HyperFirefly-b') that uses more than 75x less memory and achieves 0.03\% improvement in mean absolute accuracy.  For always-on audio classification we have shown this approach of incorporating on-device energy usage into the NAS reward function through a weighted sum is successful at optimizing the combination of energy efficiency, accuracy and memory usage of always-on models.

We believe this approach is likely general enough to be transferable to other domains (e.g. portable biomedical devices, video processing, sensor fusion). We believe the need for finding machine learning applications where for a given accuracy the energy and memory usage is as small as possible (e.g. Pareto optimal) is important to allow for larger and better networks to be deployed on lightweight batteries and small footprint and cheap SRAM chips. The use of Vizier in this study significantly removes software barriers to NAS adoption and we advocate its use in future studies due to ease of use of the API.






\subsection{Acknowledgements}
We thank our colleagues: Chansoo Lee, Hassan Rom, Kevin Kilgour, Marco Tagliasacchi, Mathieu Parvaix, Dan Ellis, Gabriel Bender, Quoc V. Le, Pete Warden, Merve Kaya, Grace Chu and Jason Rugolo for their advice.

\section{Declarations}

\subsection{Funding}
This research was funded by X, The Moonshot Factory and Google. The authors of this article were all employed at either X or Google during the time the research was carried out.
\subsection{Conflicts of interest/Competing interests}
The authors declare no competing interests.
\subsection{Ethics approval}
Not applicable. No human subjects or animals were used in this study. As such, no ethics approval was sought after.
\subsection{Consent to participate}
Not applicable.
\subsection{Consent for publication}
Not applicable. The authors do not publish any individual's data or images in this article and use previously available open source data.
\subsection{Availability of data and material}
The AudioSet dataset used to train models in this paper is openly accessible and can be found at this url: \url{https://research.google.com/audioset/}.
\subsection{Code availability - (software application or custom code)}
The neural architecture search was performed using Google's Vizier API which has recently been open sourced~\citep{bib31}. More information about Vizier's algorithms and compatibility with Google Cloud Platform can also be found in the Vertex Vizier documentation~ \citep{bib32}.
\subsection{Authors' contributions}
D.S., K.M, S.C, and M.S. conceived and planned the experiments. D.S. wrote the bulk of the code to perform the investigation. K.M reviewed the bulk of the code, wrote necessary software and provided guidance on the experiments and software architecture. The NAS experiments were run by D.S. and K.M.  S.P. worked on all matters relating to Vizier such as the search algorithm choice, early stopping, writing/reviewing code, debugging, and analyzing results. T.Z. set up the experiments to gather mobile phone power data and wrote code to gather and interpret results. All authors contributed to the writing of the paper.

\bibliography{sn-bibliography}


\end{document}